\documentclass[conference]{IEEEtran}
\IEEEoverridecommandlockouts
\usepackage{cite}
\usepackage{amsmath,amssymb,amsfonts}
\usepackage{bbm}
\usepackage{algorithmic}
\usepackage{graphicx}
\usepackage{textcomp}
\usepackage{xcolor}
\usepackage{dsfont}
\usepackage{physics}
\usepackage{comment}
\usepackage{threeparttable}
\usepackage[acronym]{glossaries}
\def\BibTeX{{\rm B\kern-.05em{\sc i\kern-.025em b}\kern-.08em
    T\kern-.1667em\lower.7ex\hbox{E}\kern-.125emX}}

\usepackage{amsthm}
\usepackage{booktabs}\usepackage{glossaries}

\newacronym{qkd}{QKD}{Quantum Key Distribution}
\newacronym{wdm}{WDM}{Wavelength Division Multiplexing}
\newacronym{fwm}{FWM}{Four-Wave Mixing}

\theoremstyle{remark} 
\newtheorem{remark}{Remark}

\begin{document}

\title{Opportunistic QKD: Exploiting Idle Capacity of Classical WDM Systems
\thanks{The authors wish to thank Daniel Wagner for valuable discussions. 
Funding was received via the Q-net-Q project from the European Union’s Digital Europe Programme under grant agreement No 101091732, and by the German Federal Ministry of Research, Technology and Space (BMFTR). This work was further financed by the DFG via grant NO 1129/2-1, by the state of Bavaria via the 6GQT and the NeQuS projects and by the BMFTR via grants 16KISQ077, 16KISQ168, 16KIS1598K, 16KIS2604 and 16KISQ093. The authors acknowledge the financial support by the Federal Ministry of Research, Technology and Space of Germany in the programme of “Souverän. Digital. Vernetzt.”. Joint project 6G-life, project identification number: 16KISK002.}
}

\author{Sumit Chaudhary, Pere Munar Vallespir, Alonso Viladomat Jasso, Janis N\"otzel (Member, IEEE) \\ ~\IEEEmembership{}
        
\IEEEauthorblockA{\textit{Emmy-Noether Group Theoretical Quantum Systems Design} \\
\textit{Technical University of Munich}\\
Munich, Germany \\
\{sumit.chaudhary, pere.munar, viladomat.jasso, janis.noetzel\}@tum.de }
}

\maketitle

\begin{abstract}
While \gls{qkd} has been proven in lab environments, large-scale implementation requires integration with existing infrastructure. This paper proposes an opportunistic \gls{qkd} framework that takes advantage of idle spectral capacity, that is, unused channels in classical fibers, to perform \gls{qkd} while prioritizing classical traffic. To mitigate crosstalk during the co-propagation of classical and quantum signals, we require a guardband of unused channels between classical and quantum signals. We propose a stochastic traffic model, with a deterministic day-night cycle and fractional Gaussian noise. Monte-Carlo simulations of an 80-channel WDM system with our stochastic traffic model demonstrate that 45-65\% of unused spectrum can be repurposed for \gls{qkd}, depending on the traffic conditions. We also model a key reservoir, with available and recovery states. We define the Reliability Horizon as the $3\sigma$ depletion threshold. We find a trade-off between buffer reset levels: increasing the buffer reset level extends the reliability horizon but linearly increases recovery time, resulting in longer service "dark windows". Furthermore, simulations indicate that the first-passage time follows a heavy-tailed distribution, which is accurately characterized by a composite model combining a diurnal trend and a Bihill transition function. This framework enables network operators to optimize buffer parameters for specific Service Level Agreements (SLAs) in real-world environments.
\end{abstract}


\section{Introduction}
\gls{qkd} represents one of the most mature quantum communication technologies and is approaching practical deployment in real-world networks \cite{scarani2009security, pirandola2020advances}. To date, most field trials and early deployments have relied on dedicated optical fiber infrastructures exclusively provisioned for quantum communication \cite{peev2009secoqc, chen2021integrated, dynes2019cambridge}. While this approach ensures controlled operating conditions and minimizes interference, it entails substantial infrastructure costs, which significantly limit scalability for widespread adoption.

An alternative and more cost-effective paradigm is the integration of quantum and classical communication within the same optical fiber infrastructure. Recent experimental studies have demonstrated the feasibility of co-propagating quantum and classical signals in \gls{wdm} systems \cite{patel2012coexistence, eraerds2010quantum, kumar2015coexistence}. However, such coexistence introduces significant technical challenges due to the extreme sensitivity of quantum signals, which typically operate at or near the single-photon level. Consequently, stringent control of system parameters is required to preserve the integrity and security of the quantum channel \cite{chaudhary2024robustness}.

In particular, impairments induced by fiber nonlinearities—most notably Kerr effects \cite{chapuran2009optical} and Raman scattering \cite{qi2010feasibility}—play a dominant role in degrading quantum signal performance. These effects necessitate careful spectral engineering, including sufficient wavelength separation between classical data channels and quantum channels, as well as power management strategies for classical signals. Furthermore, conventional WDM systems are highly optimized for classical data transmission, and incorporating quantum channels imposes additional constraints on network operation and resource allocation \cite{ware2025quantum, gagliano2024quantum, wang2017long, gavignet2023co}.

Interestingly, modern optical networks are typically provisioned with capacities that reliably handle peak traffic but significantly exceed average traffic demands, and network utilization exhibits pronounced temporal variations, including strong diurnal patterns \cite{wlodarczyk2024machine, huang2007spare, drakos2012performance}. This average underutilization presents an opportunity to opportunistically allocate unused \gls{wdm} channels for \gls{qkd} without compromising classical service quality. In this context, quantum key generation can be dynamically scheduled during periods of low classical traffic, while maintaining strict priority for classical communication.

In similar context, previous work has addressed the concept of \emph{generating entanglement while idle (GEWI)}, modelling a quantum link which prioritizes data traffic with opportunistic entanglement assistance for data transfer over entanglement generation \cite{gewi}. This methodology was investigated by integrating quantum simulators, such as QuNetSim, with network emulation environments like ComNetsEmu, thereby enabling the study of near-term protocols that exploit stored entanglement to enhance data transmission performance. \cite{diadamo2021integrating}.

Later work \cite{besser2023reliability} studied the special case of secret-key budget from a statistical perspective to analyze the fundamental trade-offs between system reliability and transmission latency.

In this work, we demonstrate that, under realistic traffic conditions, sufficient transmission opportunities exist to generate substantial quantum key material. These keys can be accumulated in buffer pools and subsequently used for securing latency-sensitive and mission-critical communications. Moreover, we analyze the temporal dynamics of the quantum key buffer and identify a fundamental trade-off between the duration of uninterrupted service—defined as the reliability horizon—and the subsequent recovery latency. By modeling the buffer as a stochastic regenerative process subject to diurnal trends and long-range-dependent (LRD) classical traffic, we characterize the first-passage time to depletion. This framework provides a robust method for network operators to optimize buffer reset levels in alignment with specific Service Level Agreements (SLAs), ensuring high-availability, quantum-secure communication across various traffic conditions.

The remainder of this paper is organized as follows. Section \ref{Classical Data Traffic Modelling} details the stochastic modeling of classical data traffic in optical fibers, incorporating diurnal trends and long-range dependence. Section \ref{Quantum-Classical Coexistence} presents the framework for quantum and classical channel coexistence in WDM systems and develops the dynamics of the opportunistic key buffer. Section \ref{sec:results} evaluates the simulation results concerning the reliability horizon, recovery latency, and first-passage time to depletion. Finally, Section \ref{conclusion} summarizes the key findings and provides concluding remarks.

\section{Classical Data Traffic Modeling}
\label{Classical Data Traffic Modelling}

The global internet infrastructure relies predominantly on optical fiber as the physical backbone for long-haul and metro-area data transmission. To achieve high aggregate throughput, these systems employ \gls{wdm}, which partitions the available optical bandwidth into multiple independent channels operating at distinct carrier frequencies. 

The stochastic nature of internet traffic introduces significant variability across multiple timescales, ranging from millisecond-level packet bursts to diurnal cycles. Mathematically, while early models approximated traffic using Poisson distributions, contemporary data traffic often exhibits self-similarity and long-range dependence (LRD), characterized by high peak-to-average power ratios (PAPR). To maintain a high Quality of Service (QoS) and prevent packet loss during transient congestion, optical links are typically overprovisioned. 

Empirical data indicate a profound underutilization of existing fiber infrastructure \cite{winzer2018fiber}. A substantial portion of the underground fiber plant remains unlit (inactive), serving as a strategic reserve to accommodate long-term, exponential demand growth without incurring additional civil engineering costs. These unused fibers could be used for \gls{qkd} until classical data demands require their use.

Even within active (lit) fibers, the average traffic load remains well below the theoretical maximum capacity. This "headroom" is a deliberate design choice necessitated by the bursty nature of traffic. Operators maintain a buffer to handle sudden traffic spikes that would otherwise cause service crashes. Consequently, rather than being fully saturated, the available \gls{wdm} channels within a typical optical fiber operate with a significant margin of idle capacity. This systemic underutilization presents both a challenge for cost-efficiency and an opportunity for dynamic resource allocation.

\subsection{Traffic Characterization and Stochastic Modeling}
Statistical analysis of Internet traffic shows that the following properties are typical of a link under our assumptions 
\begin{itemize}
    \item Diurnal Periodicity: Traffic volume exhibits a day-night cycle, as noted in \cite{sha_2022}. This periodicity is highly correlated with local activity in terrestrial links; however, in transcontinental submarine cables, these oscillations are often dampened due to the aggregation of multiple time zones.
    \item Spectral Composition: Frequency-domain analysis identifies a dominant peak at $f = 1/24$ h$^{-1}$, with secondary harmonics at integral multiples of the diurnal frequency. Furthermore, a clear "weekend effect" is observed, in which aggregate demand declines significantly relative to weekday peaks \cite{rzym2020time}.
    \item Autocorrelation and Burstiness: Internet traffic demonstrates significant temporal autocorrelation, where future arrival rates depend statistically on past states \cite{sha_2022, goscien2021modeling}. This manifests as "bursty" behavior, characterized by rapid, high-amplitude fluctuations across various timescales.
    \item Statistical Distribution: Recent dataset analyses suggest that the probability density function (PDF) of internet traffic flows is most accurately characterized by a log-normal distribution, rather than a traditional Gaussian or Poisson distribution \cite{antoniou2002log,alasmar2019distribution}.
\end{itemize}
Our study is based on the statistical properties of a link connecting two nodes in similar time zones, used for general user traffic, and we present the potential parameter values for the link in different scenarios in Table~(\ref{tab:regimes}).
To encapsulate these properties, we model the instantaneous traffic rate $R(t)$ as a modulated stochastic process:
\begin{equation}
    R(t) = R_0\, m(t) \nu(t)
\end{equation}
where $R_0$ is a constant,  maximum total deterministic traffic contribution, $m(t)$ represents the periodic deterministic trend evolution and $\nu(t) =\exp(\sigma X_H(t))$ constitutes the stochastic fluctuations of the traffic. 

\subsection{Deterministic and Stochastic Formulation}
Given the dominance of the diurnal cycle, we choose the deterministic trend evolution $m(t)$ to be sinusoidal with a period of 1 day:
\begin{equation}
    m(t) =  \left[1 - \alpha \sin^2\left(\frac{\pi t}{T_{24}}\right)\right]
\end{equation}
Here, $\alpha \in [0,1]$ denotes the amplitude of daily oscillations (representing the ratio of peak-to-trough demand), and $T_{24} = 24$ hours is the fundamental period. 

Statistical fluctuations $\nu(t)$ are modeled with an exponential of a Fractional Gaussian Noise (fGn) process, $X_H(t)$, scaled by the volatility factor $\sigma$. $X_H(t)$ is characterized by the Hurst parameter $H \in (0.5, 1)$, which governs the degree of persistence in the traffic. Thus, the noise is given by $\exp(\sigma X_H(t))$. This formulation captures the following properties of traffic:

\begin{itemize}
    \item Multiplicative Nature and Positivity: Unlike additive models, the exponential formulation inherently ensures $R(t) > 0$, consistent with physical data rates. Furthermore, it accounts for the heteroscedasticity observed in real-world networks, where absolute volatility scales proportionally with the mean traffic volume.
    \item Long-Range Dependence (LRD): By incorporating fGn, the model captures LRD \cite{leland2002self}. This is vital for representing the "heavy-tailed" nature of modern traffic and the persistence of congestion bursts, phenomena that traditional Markovian models fail to replicate.
    \item Cyclostationarity and Resource Harvesting: The periodic term reflects the physical reality of \gls{wdm} backbone links. By identifying periods of low utilization (nighttime troughs), the model quantifies the "idle capacity" available for opportunistic secondary services, such as \gls{qkd} or low-priority background data transfers.
\end{itemize}

While internet traffic is also known to exhibit self-similarity \cite{leland2002self}, where statistical properties remain invariant across disparate time scales, this property is mainly relevant for packet-level queuing analysis. These models use Fractional Auto-Regressive Integrated Moving Average (FARIMA) \cite{adas2002traffic}. Since our objective is to quantify idle capacity with lower temporal resolution, focusing on how many resources are available without studying single-packet behavior, we use a simpler model that lacks self-similarity, as we assume, for our application, that such behavior would not make a significant contribution. Small fluctuations would average out at the timescales considered here. 

\begin{figure}
    \centering
    \includegraphics[width=1\linewidth]{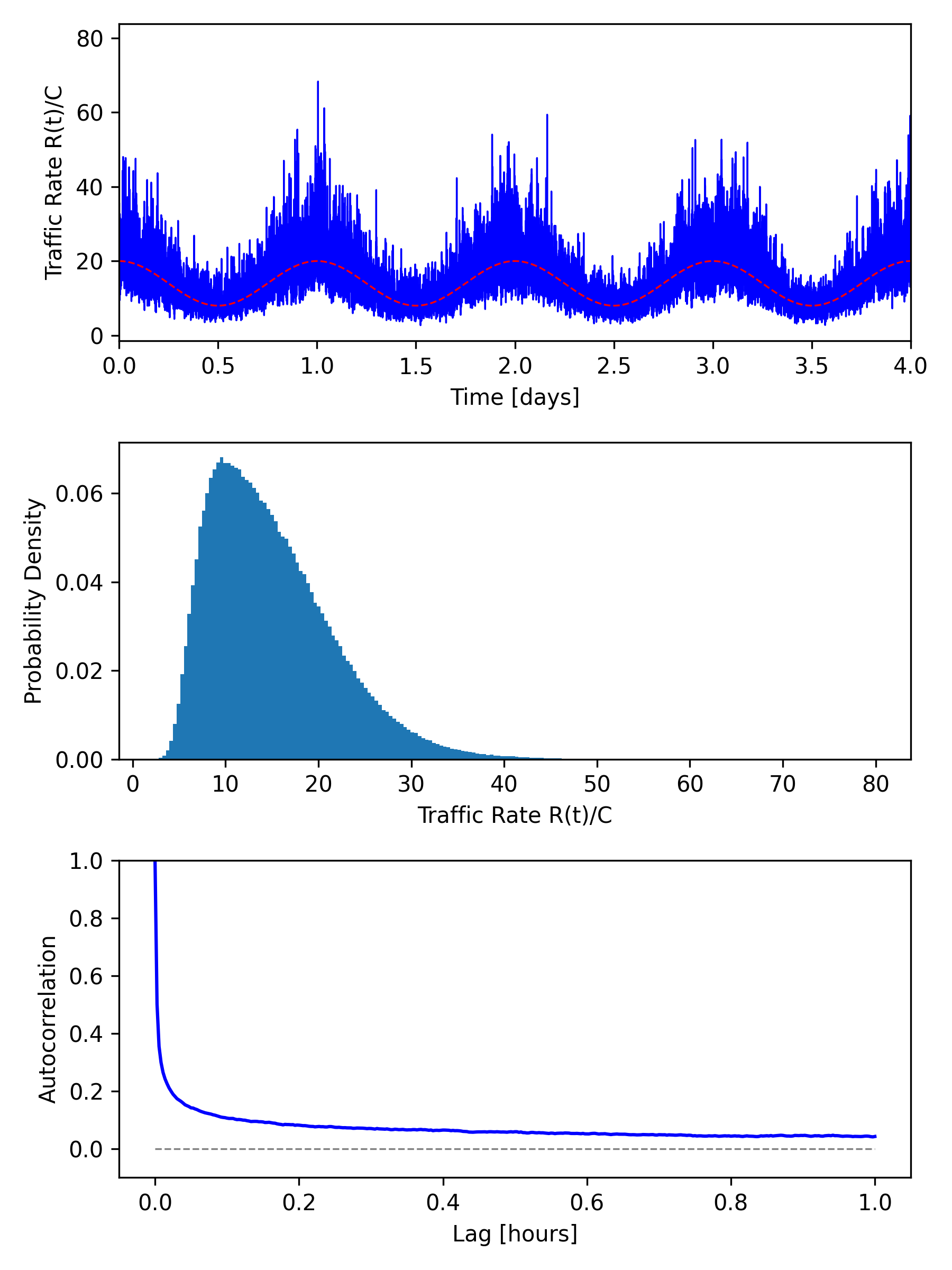}
    \caption{Representative stochastic realization of classical data traffic modeled with parameters $\alpha=0.6$, $H=0.8$, $\sigma=0.3$ and $p=20$. The profile demonstrates the diurnal trend, heavy-tail distribution, and long-range dependence through autocorrelation. ($\alpha$ is set to zero for the autocorrelation plot to avoid periodic correlations from the diurnal trend)}
    \label{fig: example traffic}
\end{figure}

We show an example of the traffic values produced by our model in Figure~(\ref{fig: example traffic}). As expected, traffic exhibits strong fluctuations during high traffic and shows strong autocorrelation on the scale of 10-30 min. 

\subsection{Statistical Moments of Traffic Rate}
To analyze the buffer dynamics and key consumption, we derive the instantaneous mean and variance of $R(t)$. Given that $X_H(t) \sim \mathcal{N}(0, 1)$, the stochastic component $Z(t) = \exp(\sigma X_H(t))$ follows a log-normal distribution.

The mean traffic rate at time $t$ is given by:
\begin{equation}
    \mathbb{E}[R(t)] = R_0\ m(t)\ e^{\frac{\sigma^2}{2}}
    \label{mean_ln}
\end{equation}

The variance of the traffic rate is given by 
\begin{equation}
    \mathbb{V}[R(t)] = \left[ R_0\ m(t)   \right]^2 e^{\sigma^2} \left( e^{\sigma^2} - 1 \right)
    \label{variance_ln}
\end{equation}
where equations \ref{mean_ln} and \ref{variance_ln} follow directly from the definition of the first and second moments of a log-normal random variable.
 In particular, the variance depends only on time through deterministic fluctuations, as the network experiences larger absolute fluctuations during peak hours. This statistical property is relevant for choosing the size of the \gls{qkd} key buffer, as the probability of "key starvation" will be significantly higher during the daily cycle's maxima.

\section{Quantum-Classical Resource Coexistence in WDM Systems}
\label{Quantum-Classical Coexistence}
We introduced a model for the classical data traffic in the previous section. In the following, we show how channels are assigned to \gls{qkd} and classical data transmission depending on traffic. 

Consider a \gls{wdm} link comprising $N$ discrete channels, each with a classical transmission capacity $C$. At any time $t$, the number of channels required to support the total classical traffic demand $R(t)$ is given by:
\begin{equation}
    N_C(t) = \left\lceil \frac{R(t)}{C} \right\rceil = \left\lceil p \  m(t) \nu(t) \right\rceil
\end{equation}
where $p = R_0/C$ represents the peak-load-to-channel-capacity ratio. 

In a co-propagation environment, \gls{qkd} performance is strictly limited by nonlinear effects, most notably \gls{fwm} and Raman scattering, caused by the high-power classical channels \cite{peters2009dense}. Within the WDM framework, we consider $N_C$ classical channels operating on a frequency grid defined by $f_i = f_0 + i\Delta f$, where $i \in \{0, 1, \dots, N_C-1\}$ and $\Delta f$ represents the channel spacing. Nonlinear interactions via \gls{fwm} generate parasitic frequencies at $f_{ijk} = f_i + f_j - f_k$ (for $i, j \neq k$). When the primary $N_C$ channels are occupied by high-power classical signals, the resulting FWM products manifest as significant noise across the subsequent $N_C-1$ spectral slots. To mitigate this crosstalk, we adopt a conservative allocation strategy: the presence of $N_C$ active classical channels necessitates a guard band of the next $N_C-1$ adjacent spectral slots, rendering them unusable for quantum key distribution.

Then, the number of available \gls{qkd} channels, $N_Q(t)$, is given by
\begin{equation}
    N_Q(t) = N - 2N_C(t) + 1.
\end{equation}

By applying the fundamental bounds of the ceiling function, $x \le \lceil x \rceil < x + 1$, we derive the upper and lower analytical bounds for the available quantum resources:
\begin{equation}
    N_Q^{\pm}(t) = N \pm 1 - 2\ p\ m(t) \exp\left( \sigma X_H(t) \right).
\end{equation}

Furthermore, since $\exp(\sigma X_H(t))$ follows a log-normal distribution with a mean of $e^{\sigma^2/2}$, the expected availability of quantum channels over time can be expressed as:
\begin{equation}
    \mathbb{E}[N_Q^{\pm}(t)] = N \pm 1 - 2\ p\ m(t)\ e^{\frac{\sigma^2}{2}}.
\end{equation}

\subsection{Network Traffic Regimes}
To evaluate the system's performance across diverse networking environments, we define three distinct traffic categories. These categories represent fundamental shifts in how the physical layer must adapt to varying classical loads and volatility:

\begin{table}[htbp]
\centering
\begin{threeparttable} 
\caption{Simulation Parameters for Different Traffic Categories}
\label{tab:regimes}
\begin{tabular}{@{}lccccc@{}}
\toprule
\textbf{Traffic type} & \textbf{Volatility} & \textbf{Diurnal fluctuation} & \textbf{$p/N$} & \textbf{$\alpha$} & \textbf{$\sigma$} \\ \midrule
Category 1                 & Medium              & High                 & 0.5            & 0.875             & 0.3               \\
Category 2                 & Low                 & Low                  & 0.2            & 0.15              & 0.08              \\
Category 3                 & High                & Medium               & 0.2            & 0.6               & 0.8               \\ \bottomrule
\end{tabular}
\begin{tablenotes} 
    \item[] Note: The parameter values listed are for illustrative purposes within the considered internet traffic model.
\end{tablenotes}
\end{threeparttable}
\end{table}

Each traffic regime in Table~(\ref{tab:regimes}) has an intuitive explanation.

\begin{enumerate} 
    \item Category 1 : ($p/N = 0.5, \alpha=0.875, \sigma=0.3$) \\
    \textit{Physical Interpretation:} This regime is typical of enterprise or campus backbones. The high modulation depth ($\alpha$) indicates that resource availability is governed almost entirely by the predictable 24-hour socio-economic cycle. The moderate volatility ($\sigma$) suggests that while some noise exists, the primary driver is the nighttime trough, allowing for scheduled, high-bandwidth quantum tasks during off-peak hours.
    
    \item Category 2 : ($p/N = 0.2, \alpha=0.15, \sigma=0.08$) Characteristic of Machine-to-Machine (M2M) or IoT-heavy data flows. Here, the minimal diurnal variation and exceptionally low volatility ($\sigma$) provide a quasi-static environment. This stability is ideal for continuous, steady-state QKD key generation, as the number of available quantum channels remains highly predictable with minimal interruptions. 

    \item Category 3 : ($p/N = 0.2, \alpha=0.6, \sigma=0.8$) \\ Representative of edge or residential access networks. In this scenario, despite a low average occupancy ($p/N$), the high volatility ($\sigma$) and stochastic bursts present the primary challenge to coexistence. The rapid fluctuations in classical demand make quantum channel availability highly unpredictable, necessitating the use of elastic or adaptive QKD protocols to handle the high-frequency switching of available spectral slots.
\end{enumerate}

\begin{remark}[Saturated Regimes and Quantum Outage]
It is important to note that our analysis focuses on operational regimes where $p/N \le 0.5$. In scenarios characterized by saturated traffic ($p/N > 0.5$) combined with low modulation ($\alpha \ll 1$), classical data occupancy occupies a dominant portion of the WDM spectral window. Under these conditions, the stringent guard-band requirements necessary to mitigate nonlinear crosstalk on \gls{qkd} cannot be met without exceeding the fiber's physical capacity or forcefully dropping/delaying part of the classical traffic. 

Such high-load states result in an outage of \gls{qkd}-generated keys, as the high density of classical signals physically prevents the fiber from being used for \gls{qkd}. Because these saturated states lead to no idle channels, they are excluded from the current performance analysis. However, this can occur temporarily in our models when traffic undergoes large stochastic fluctuations. 
\end{remark}

\begin{figure}
    \centering
    \includegraphics[width=1\linewidth]{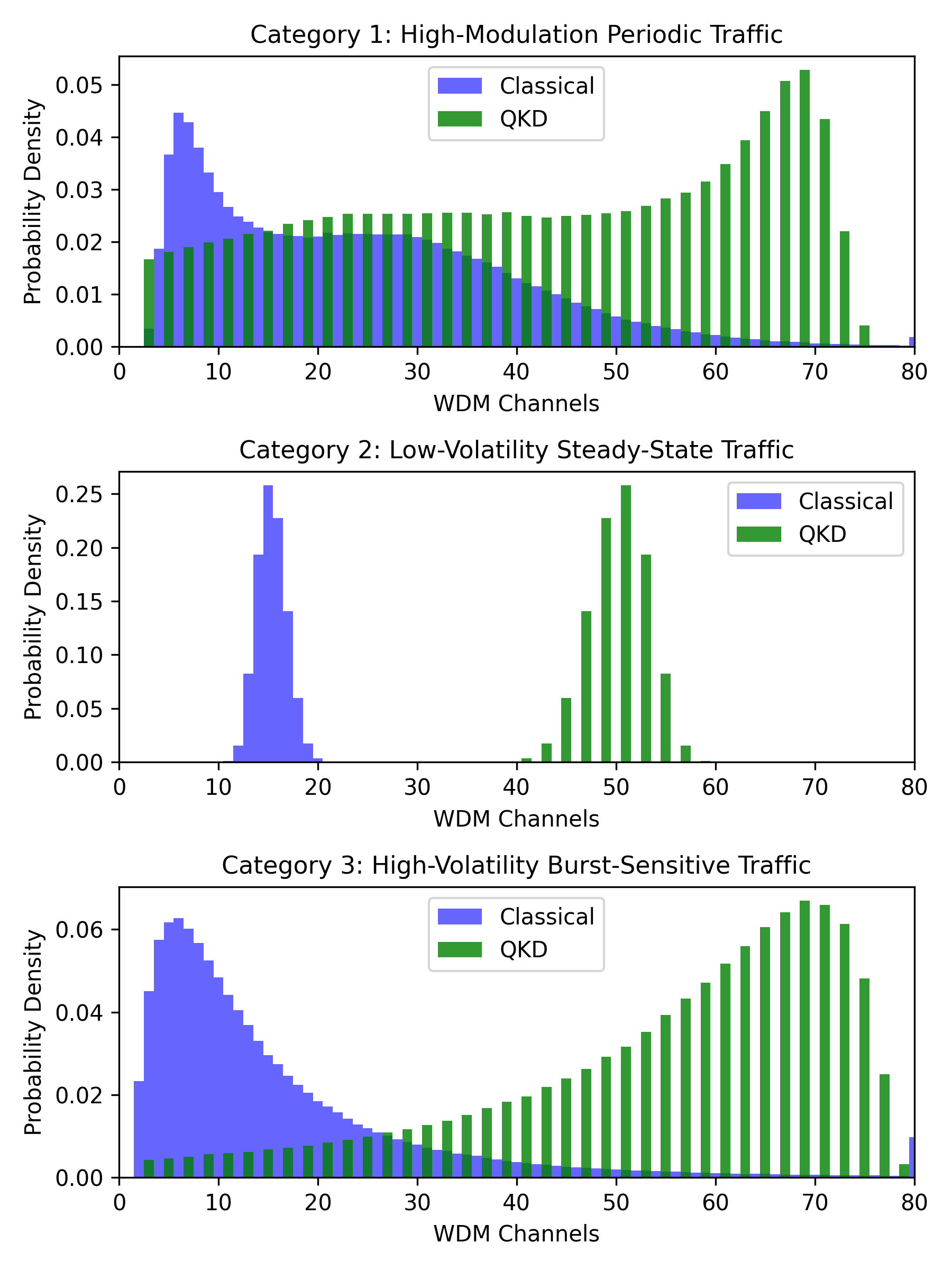}
    \caption{Distribution of classical WDM channel occupancy (blue) and the corresponding opportunistic QKD channel availability (green). The distributions are evaluated across three distinct traffic categories, illustrating the inverse relationship between classical load and available spectral resources for \gls{qkd}. } 
    \label{fig_traffic_regimes}
\end{figure}

\subsection{Stochastic Dynamics of the Quantum Key Buffer}

To quantify the practical utility of opportunistic resource harvesting, we model the temporal evolution of a quantum key buffer, $B(t)$, which serves as a reservoir for keys generated during periods of classical underutilization. Assuming a single-channel key rate of $C_Q$ (bits per second), the aggregate replenishment rate is directly proportional to the instantaneous number of active quantum nodes $N_Q(t)$. Assuming a consumption rate $S(t)$, the governing differential equation for the buffer state is:
\begin{equation}
    \frac{d B(t)}{dt} = N_Q(t) C_Q - S(t), \quad B(t) \ge 0.
\end{equation}

The long-term stability of the buffer is dictated by the drift coefficient. If $C_Q \mathbb{E}[N_Q(t)] > \mathbb{E}[S(t)]$, the system exhibits a positive drift, leading to buffer saturation (overflow) where surplus keys must be discarded or stored in secondary high-latency media. Conversely, if $C_Q \mathbb{E}[N_Q(t)] < \mathbb{E}[S(t)]$, the negative drift leads to rapid depletion, resulting in a service outage where classical data cannot be encrypted. We therefore focus our analysis on the balanced regime, where $C_Q \mathbb{E}[N_Q(t)] \approx \mathbb{E}[S(t)]$, as this threshold represents the maximum sustainable secure throughput of the link.

The consumption profile $S(t)$ is determined by the specific security requirements of the underlying applications. In the following, we assume constant consumption. Under this assumption, $S(t) = S_0$. This model is appropriate for applications that require a deterministic, continuous supply of key over short intervals. Typical use cases include secure voice-over-IP (VoIP) links, real-time sensor telemetry for critical infrastructure, users accessing banking, or steady-state heartbeat signals in military-grade command-and-control (C2) systems. In these scenarios, the buffer ensures these applications don't run out of keys when key production fluctuates.

\subsection{Constant consumption rate}
Under a constant consumption rate, the buffer follows the following differential equation
\begin{equation}
    \frac{d{B}(t)}{dt}=K(t)\ (e^{\frac{\sigma^2}{2}}-e^{\sigma X_H(t)})
\end{equation}
where $K(t) = 2p C_Q m(t)$  represents the deterministic modulation of the channel availability. Note that we chose the consumption so that $\mathbb{E}[\frac{dB}{dt}] = 0$, as discussed earlier. A zero-drift regime is particularly relevant for assessing system stability. 
To characterize the stability of the quantum key buffer, we derive the first and second moments of the buffer deviation process $\bar{B}(t)$. 

\subsubsection{Variance and Volatility Scaling}
The variance $\text{Var}(\bar{B}(t))$ quantifies the risk of buffer depletion or overflow due to stochastic bursts. Utilizing the covariance properties of log-normal processes, the variance is expressed as:
\begin{equation}
    \sigma^2_{\bar{B}}(t) = e^{\sigma^2} \int_0^t \int_0^t K(s) K(u) \left( e^{\sigma^2 \gamma_H(s-u)} - 1 \right) ds du
\end{equation}
where $\gamma_H(\tau) = \frac{1}{2} (|\tau+1|^{2H} - 2|\tau|^{2H} + |\tau-1|^{2H})$ is the autocovariance of the fractional Gaussian noise. 
 For $H > 0.5$, the long-range dependence of the traffic causes the variance to scale as $t^{2H}$, which is faster than the linear growth ($t^1$) observed in standard Brownian motion. This implies that the "security margin" required in the buffer must grow super-linearly with time to maintain the same reliability against traffic bursts in the High-Volatility Bursty Regime.

\subsection{Buffer States and Transitions} The key buffer dynamics are modeled as a process comprising two states: available and recovery. 

\subsubsection{Available state} The available state begins with the buffer at its reset level $B_0$. During this phase, the system performs opportunistic key generation while satisfying a constant demand. To quantify the system's resilience, we define the Reliability Horizon ($t_{0}$) as the temporal boundary where the $3\sigma$ lower excursion of the stochastic buffer process reaches zero. Mathematically, $t_0$ satisfies:$$\mathbb{E}[B(t_0)] - 3\sqrt{\text{Var}(B(t_0))} = 0$$where $B(0) = B_0$. This horizon represents the duration for which the QKD system can maintain uninterrupted service with a confidence level of $99.73\%$, accounting for both diurnal classical traffic and long-range-dependent (LRD) traffic fluctuations. To facilitate a hardware-independent analysis, we normalize the buffer capacity using the Daily Key Unit (DKU). We define $1~\text{DKU} = C_Q T_{24}$ as the total expected key volume generated by a single WDM channel over a standard 24-hour cycle. This normalization allows the buffer dynamics to be analyzed independently of the specific hardware key generation rate $C_Q$. 

\subsubsection{Recovery state} The recovery state is triggered the moment the buffer level hits zero. During this interval, the \gls{qkd} service is suspended and key consumption is halted ($S(t) = 0$), rendering the service unavailable to end users. The buffer is recharged exclusively through opportunistic generation until it returns to the reset level $B_0$. The recovery duration $\tau_{rec}$ is the time required to satisfy:
\begin{equation}B_0 = \int_{t'}^{t' + \tau_{rec}^\pm} C_Q \left[ N \mp 1 - 2p \cdot m(t) \exp\left( \sigma X_H(t) \right) \right] , dt
\end{equation}
Assuming the recharge process is rapid relative to the diurnal cycle ($\tau_{rec} \ll T_{24}$), the demand function $m(t)$ can be treated as quasi-static, fixed at the moment of depletion $t'$. The expected recovery time $\mathbb{E}[\tau_{rec}]$ is thus approximated by:
\begin{equation}\mathbb{E}[\tau_{rec}^\pm] \approx \frac{B_{0}}{C_Q \left[ (N \mp 1) - 2p m(t') e^{\sigma^2/2} \right]}
\end{equation}
where the term $e^{\sigma^2/2}$ accounts for the mean of the log-normal scaling factor associated with the fractional Gaussian noise.

\section{Results and Discussion}
\label{sec:results}

In this section, we evaluate the performance of the proposed opportunistic \gls{qkd} framework. We analyze the availability of idle \gls{wdm} capacity across the previously introduced traffic categories and examine the resulting buffer dynamics, specifically, the trade-off between service reliability and recovery latency. 

\subsection{WDM Channel Availability for QKD}
Our analysis considers a fiber link with a total capacity of 80 WDM channels. By monitoring the gaps in classical communication, the system identifies idle windows for opportunistic quantum key generation. The average number of available \gls{qkd}-capable channels was calculated for three distinct traffic categories, as summarized in Table~(\ref{tab:wdm_availability}).

The statistical distribution of the available quantum channels $N_Q(t)$ across the three defined traffic regimes is illustrated in Fig. (\ref{fig_traffic_regimes}). In category 1, there is an even distribution of channels. For category 2, both transmission types exhibit less variability in the number of channels used. Category 3 exhibits intermediate behavior between categories 1 and 2.
\begin{table}[htbp]
\centering
\caption{Average QKD Channel Availability across Traffic Categories (Total Channels = 80)}
\label{tab:wdm_availability}
\begin{tabular}{@{}lccc@{}}
\toprule
\textbf{Traffic Category} & \textbf{Avg. QKD Channels} & \textbf{Utilization (\%)} \\ \midrule
Category 1                & 36.22                      & 45.28\%                   \\
Category 2                & 50.25                      & 62.81\%                   \\
Category 3                & 51.81                      & 64.76\%                   \\ \bottomrule
\end{tabular}
\end{table}

The results indicate that even under moderate classical loads (Category 1), over 45\% of the WDM spectrum can be repurposed for quantum key distribution. This provides a significant cryptographic resource capable of sustaining high-bitrate encryption for sensitive network applications.

\subsection{Reliability Horizon and Volatility Impact}
\begin{figure}[htbp]
    \centering
    \includegraphics[width=1.0\linewidth]{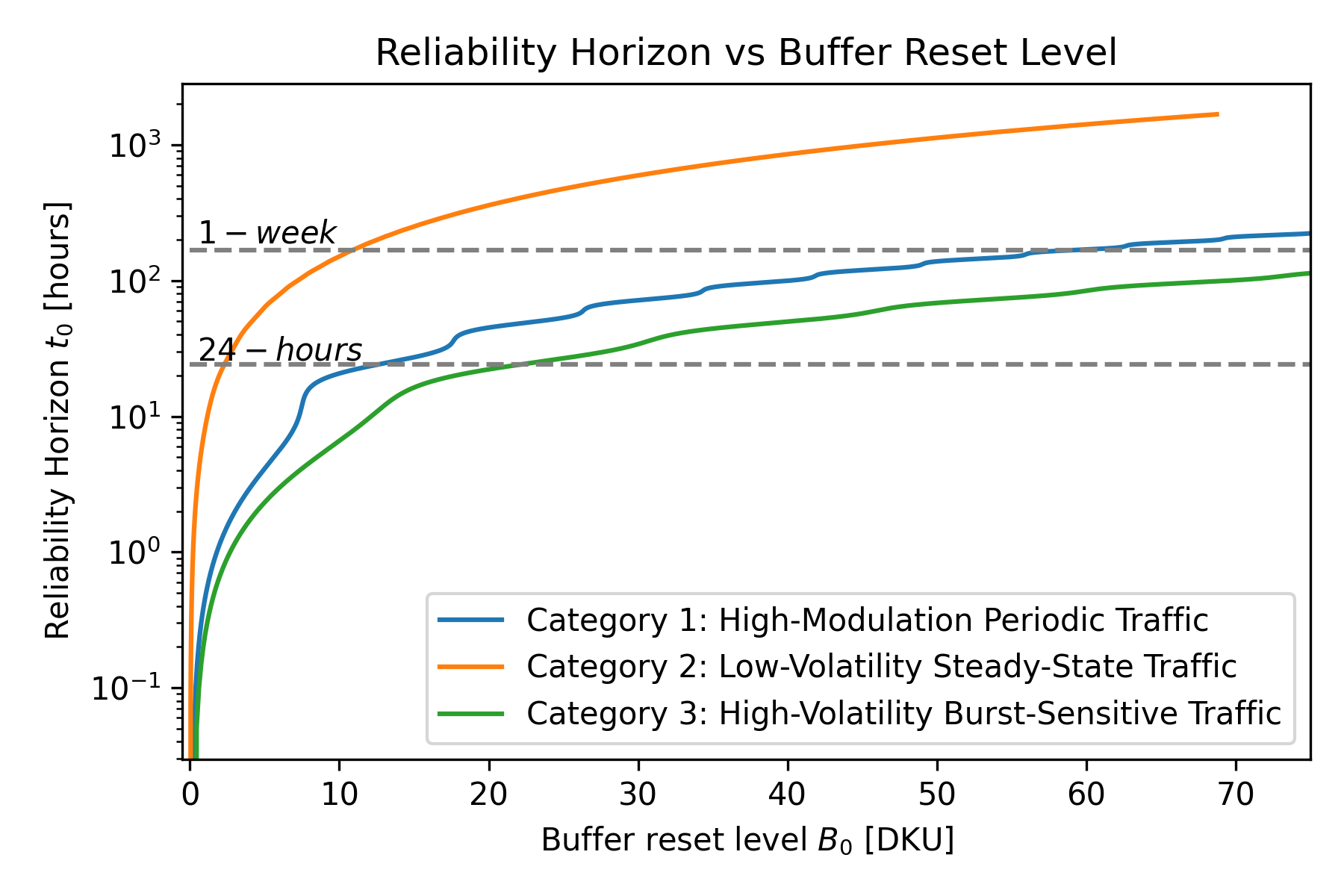}
    \caption{Reliability Horizon ($t_0$) as a function of Buffer Reset Level ($B_0$) for different volatility regimes.}
    \label{fig:reliability_horizon}
\end{figure}

The primary metric for the key reservoir's performance is the \textit{Reliability Horizon} ($t_0$). Figure~(\ref{fig:reliability_horizon}) illustrates the relationship between the buffer reset level ($B_0$) and $t_0$. As expected, a higher reset level provides a larger safety margin, leading to a longer duration of uninterrupted service.

However, the reliability of the system is highly sensitive to the volatility of the underlying classical traffic. In high-modulation and high-volatility scenarios, the reliability horizon contracts significantly compared to steady-state conditions. This is mathematically attributed to the accelerated growth of the buffer variance $\sigma^2_B(t)$. In bursty traffic regimes (characterized by higher Hurst parameters and $\sigma$ values), the increased fluctuations raise the probability of a $3\sigma$ lower excursion hitting the zero boundary, necessitating an earlier transition to the recovery state.

\subsection{Recovery Dynamics and Optimization Trade-offs}
\begin{figure}[htbp]
    \centering
    \includegraphics[width=1.0\linewidth]{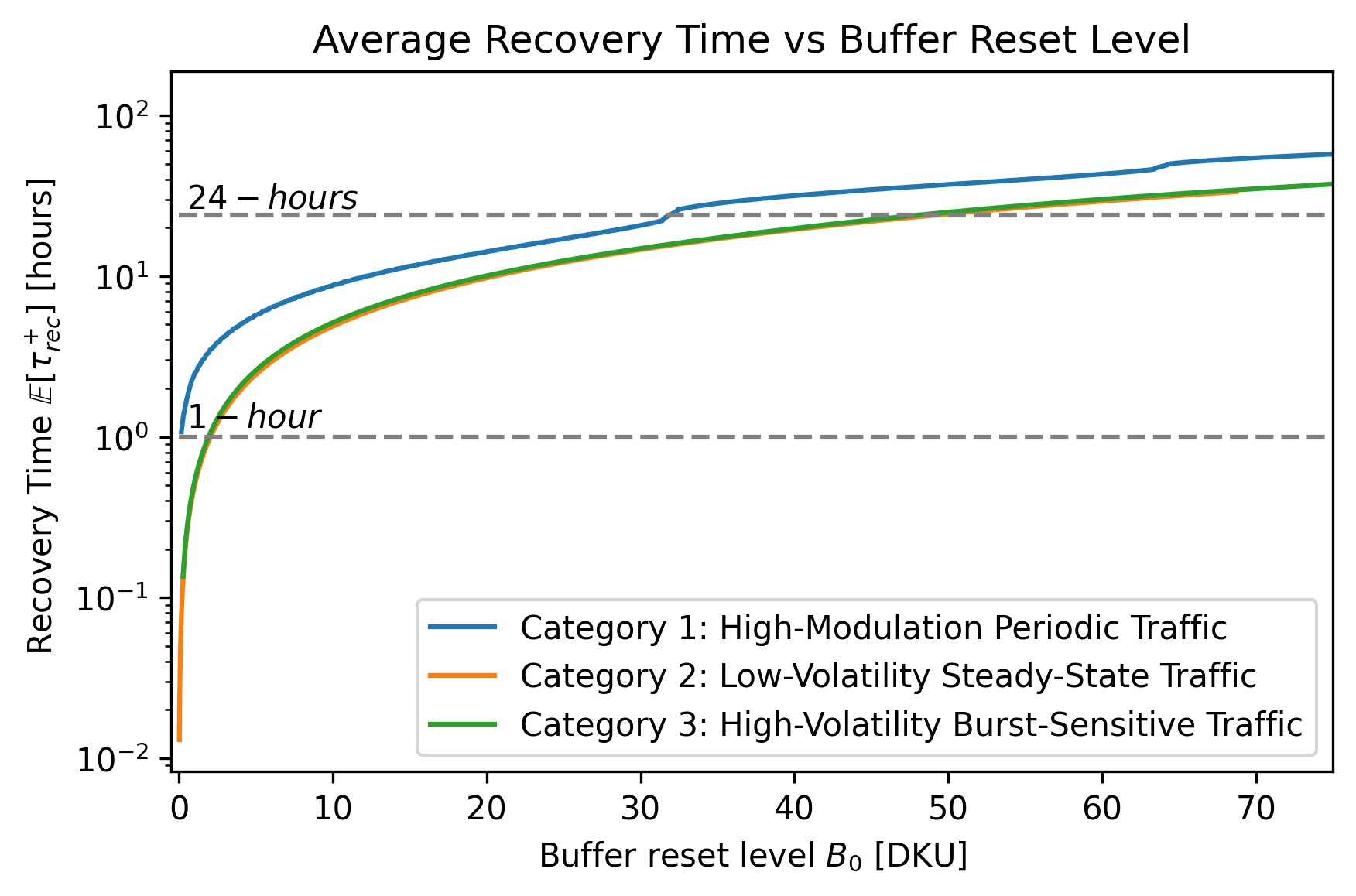}
    \caption{Expected Recovery Time ($\tau_{rec}$) vs. Buffer Reset Level ($B_0$).}
    \label{fig:recovery_time}
\end{figure}

While a larger $B_0$ extends the period of availability, it simultaneously increases the operational penalty during the recovery phase. Figure~(\ref{fig:recovery_time}) demonstrates that the expected recovery time ($\mathbb{E}[\tau_{rec}]$) scales linearly with the target reset level.

An increased recovery time results in a larger "dark window" where the \gls{qkd} service is suspended and the link is vulnerable to key depletion. This highlights a fundamental design trade-off: 
\begin{itemize}
    \item Low $B_0$: Results in frequent but brief service interruptions.
    \item High $B_0$: Ensures long-term stability but leads to prolonged periods of unavailability once the buffer is exhausted.
\end{itemize}

Consequently, the selection of the optimal buffer reset level must align with the network's specific Service Level Agreement (SLA). For mission-critical tasks requiring high uptime, a balanced $B_0$ that minimizes the outage frequency-duration product is required.

\subsection{Probability Distribution of Hitting Times}
\label{subsec:hitting_times}

\begin{figure}[htbp]
    \centering
    \includegraphics[width=1.0\linewidth]{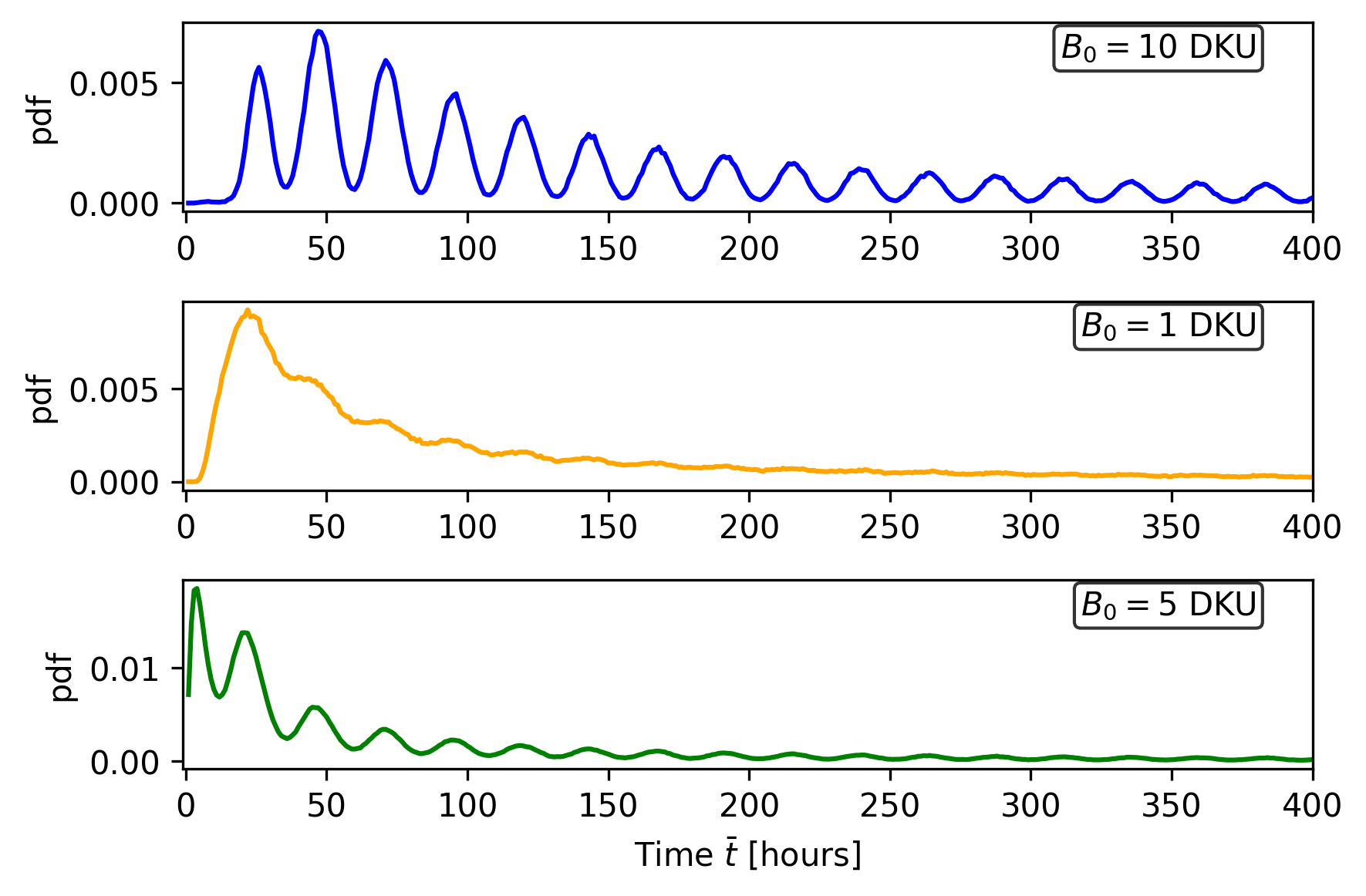}
    \caption{Probability Density Function (PDF) of the buffer hitting time $\bar{t}$ derived via Monte-Carlo simulation. For traffic categories 1, 2 and 3.}
    \label{fig:hitting_time_dist}
\end{figure}

We investigate the stochastic transition from the available to the recovery state by analyzing the \textit{First Passage Time} (FPT). Assuming the buffer begins at the reset level $B_0$ at $t=0$, we denote $\bar{t}$ as the time required for the buffer level to reach zero. The probability density function (PDF) of $\bar{t}$ is illustrated in Figure~(\ref{fig:hitting_time_dist}) for all three classical traffic categories.

 The simulated distributions reveal a pronounced heavy-tailed behavior, suggesting that while the majority of depletion events occur within a predictable timeframe, the long-range dependence ($H > 0.5$) inherent in the traffic can occasionally sustain the buffer for durations far exceeding the mean. Due to the non-Markovian nature of the fractional Gaussian noise and the presence of periodic drift, an analytical closed-form solution for this PDF is mathematically intractable. 

To provide a functional model for network resource planning, we utilized a non-linear regression to fit the simulation data. We observed that the PDF is closely approximated by the product of the diurnal trend function $m(t)$ and a \textit{Bihill transition function}, $H(t)$. This empirical relationship is expressed as:

\begin{equation}
    f(\bar{t}) \approx \mathcal{K} \cdot m(\bar{t}) \cdot H(\bar{t})
\end{equation}

where $\mathcal{K}$ is a normalization constant and $H(t)$ is the Bihill function, defined as:

\begin{equation}
    H(t) = \frac{1}{\left[ 1 + \left( \frac{a_1}{t} \right)^{m_1} \right] \left[ 1 + \left( \frac{t}{a_2} \right)^{m_2} \right]}
\end{equation}

In this formulation, the parameters $a_1$ and $m_1$ dictate the initial rise in hitting probability, while $a_2$ and $m_2$ characterize the decay rate of the distribution's tail. This composite fitting method effectively captures the interaction between the deterministic 24-hour traffic cycle and the stochastic decay of the key reservoir, offering a practical tool for predicting system outages in opportunistic \gls{qkd} deployments.

\section{Discussion and Conclusion}
\label{conclusion}
This work presents a comprehensive framework for integrating opportunistic \gls{qkd} into existing classical WDM infrastructures. We proposed an architecture that leverages the idle spectral capacity of classical channels and allocates a strict guard band of $(N_C - 1)$ channels to mitigate \gls{fwm} interference, demonstrating that a substantial portion of the \gls{wdm} spectrum can be used for quantum key generation depending on current traffic. We modeled the dynamics of the key buffer using Monte-Carlo simulations, which revealed a fundamental operational trade-off. Increasing the buffer reset level ($B_0$) significantly increases the Reliability Horizon ($t_0$), providing 99.73\% confidence in service continuity, but also results in a linear increase in recovery time ($\tau_{rec}$). Thus, network operators would need to calibrate buffer parameters according to specific Service Level Agreements (SLAs), balancing outage frequency against the duration of service interruptions. These simulations also suggested that $\bar{t}$ (the first time the buffer hits 0) has a heavy-tailed distribution which can be approximately described by a composite model combining the diurnal traffic trend with a Bihill transition function. 

While the guard band rule used in this work provides a valuable benchmark, it is inherently simplistic as it primarily isolates \gls{fwm} as the dominant impairment source affecting the quantum channels. In a real-world deployment, other critical physical-layer parameters—such as the launch powers of the classical signals, inter-channel isolation within the WDM demultiplexers, and spontaneous Raman scattering (SpRS)—simultaneously degrade \gls{qkd} performance. A more rigorous, holistic impairment model is therefore required to comprehensively map the multi-dimensional parameter space in which classical and quantum communications can coexist. Nevertheless, numerous experimental studies have already demonstrated that such co-propagation is practically viable \cite{dou2024coexistence, wang2017long, gavignet2023co}. Defining the exact theoretical boundaries and system parameter thresholds of both \gls{qkd} and classical systems under which co-propagation remains sustainable remains an active, ongoing area of investigation.

Ultimately, this architecture establishes a solid foundation for a smooth transition from dedicated QKD links to cost-effective, shared-fiber quantum-secure networks. By integrating these opportunistic \gls{qkd} methods with the redundant key management protocol \cite{chaudhary2024distance} for quantum networks, the proposed system could be scaled to support robust, end-to-end key distribution across several thousand kilometers. Such a combination would allow for higher security standards and increased key rates without the strict requirement for truly trusted intermediate nodes.

Future research should investigate advanced buffer reset strategies that dynamically adapt to real-time traffic fluctuations and variable key consumption rates. Specifically, this involves exploring differentiated buffer policies that assign tiered priority access to the key buffer based on traffic classification (e.g., critical financial data versus standard encrypted web traffic), thereby optimizing the $B_0$ and $\tau_{rec}$ trade-offs per service class. Additionally, the proposed framework can be extended to study entanglement sharing and multi-party quantum protocols—such as quantum conference key agreement (QCKA)—within a shared WDM environment alongside classical traffic. Finally, experimental validation on a physical testbed is essential to evaluate real-world \gls{qkd} performance and assess the practical efficacy of the proposed classical-quantum network integration strategy.

\bibliography{ref}
\bibliographystyle{IEEEtran}

\end{document}